# Single-layer $Ga_2O_3$/graphene heterogeneous structure with optical switching effect

Lijie Li

College of Engineering, Swansea University, Bay Campus, Swansea, SA1 8EN, UK

Email: L.Li@swansea.ac.uk

**Abstract**: Both single layer $Ga_2O_3$ (SLGO) and graphene are attractive due to their respective electronic and mechanical properties such as wide bandgap and high electrical conductivity. Bringing them together by using van der Waals force to form a heterogeneous structure is new and worth to investigate. In this work, density functional theory (DFT) study of SLGO/graphene has been conducted through varying the interlayer distance. Standard procedures of calculating double-layer heterostructures using DFT has been followed, and several interesting phenomena have been unveiled, for example, band opening in conduction bands of the SLGO and graphene and switching effect of the in-plane optical absorption.

**Introduction**:

Heterostructures based on two-dimensional (2D) layered materials bonded by van der Waals (vdW) force have attracted much research interests attributed to their unique electronic, optical, and thermal properties [1] [2] [3]. Notable advancements include stacking two graphene layers for superconductivities especially when they are aligned with certain angles [4] [5]. High thermoelectric properties have been reported with twisted bi-layer graphene [6] and defected graphene heterostructure [7]. Optical properties could be tuned with layered 2D heterostructures, both demonstrated experimentally in [8] and theoretically through DFT in [9].

Ultra-wide band (4.4-4.9 eV) of $\beta$-$Ga_2O_3$, one of thermodynamically stable crystal phases, makes it suitable for high performance power electronic devices [10] [11] [12], solar-blind detector [13], and gas sensors [14]. $\beta$-$Ga_2O_3$ has a centrosymmetric crystal structure, hence lacks polarisation properties. However, recent analysis showed that substitutionally doped $\beta$-$Ga_2O_3$ can have piezoelectric properties [15]. Two-dimensional (2D) form of the $\beta$-$Ga_2O_3$ has been investigated as it offers advantages such as large surface to volume ratio and possible quantum confinement effect. It has been investigated recently both in terms of experimental realisation [16] [17] and theoretical modelling [18] [19], despite the fact that there was a wide range for the calculated bandgaps of the 2D $Ga_2O_3$. Monolayer $Ga_2O_3$ constructed similarly to α-$In_2Se_3$ crystal structures has been simulated using DFT [20]. Stacking graphene with $Ga_2O_3$ has sparked many interests as graphene exhibits high electron mobility attributed to the very low effective electron mass, which was used as the transparent conductive layer in a ultraviolet sensor [21]. Another benefit from stacking graphene and $Ga_2O_3$ to form heterostructure was that 2D materials don't have surface dangling bonds and have very weak interacts with the $Ga_2O_3$ leading to tuneable interface properties [22]. Considering that both 2D $Ga_2O_3$ and graphene offer great electronic and optical properties, stacking graphene with the single layer $Ga_2O_3$ (SLGO) will potentially give more interesting performances due to interaction of these two layers. In this work, a SLGO/graphene heterostructure has been constructed and subsequently analysed using DFT. Band opening and switching effect of the in-plane optical absorption have been unveiled for this heterostructure.

**Computational procedure and results**:

Calculations in this work were conducted using the DFT package – Quantum Atomistix ToolKit simulation tools [23]. First of all, monolayer of β-$Ga_2O_3$ has been created in the a-direction according to the schematic graph shown in Figure 1a. Optimised lattice parameters for the monolayer β-$Ga_2O_3$ are: a = 5.763 Å, b = 5.976 Å, c = 62.545 Å (including two 25 Å vacuum layers on the top and bottom of the SLGO), β = 103.648°. It contains 20 atoms (Ga8O12), and the space group is P2. Before slicing into 2D-Ga2O3, the optimised parameters of the bulk Ga2O3 cell are: a = 5.881 Å, b = 6.179 Å, c = 12.427 Å, β = 103.648°, which contains 40 atoms (Ga16O24), belonging to the space group of C2/m. Similar approach to build 2D $Ga_2O_3$ was described in a previous reference [24]. Bi-layer structure consisting of graphene and SLGO was then constructed using the generalized lattice match (GLM) procedure described in [25] with the interface toolkit of the DFT package based on the balanced criteria of minimising both the number of atoms and interface strains. After optimisation of the SLGO and graphene, the heterostructure was build using the interface builder tools embedded in the software. The two single-layer cells were dropped to the interface builder, and various bilayer settings were calculated in terms of number of atoms and mean absolute strain between these two layers using the co-incidence site lattice method. To minimise structural instability, it was intended to find small mean absolute strain while in the meantime a small number of atoms was preferred for computational power consideration. To satisfy the balance between a little mismatch and a small number of atoms, the heterostructure consisting of $C_{14}Ga_8O_{12}$ with 2.64% mean absolute strain (MAS) was chosen. It was described in previous studies the mismatch strains for various heterostructures were usually between 1% - 4% and no more than 5% [26] [27]. Here the MAS has been obtained using the strain to both surfaces equally, then geometrical optimisation has been conducted to relax the heterostructure before property calculations. The atomic model of the SLGO/graphene bilayer structure is shown in Figures 1b-1d.

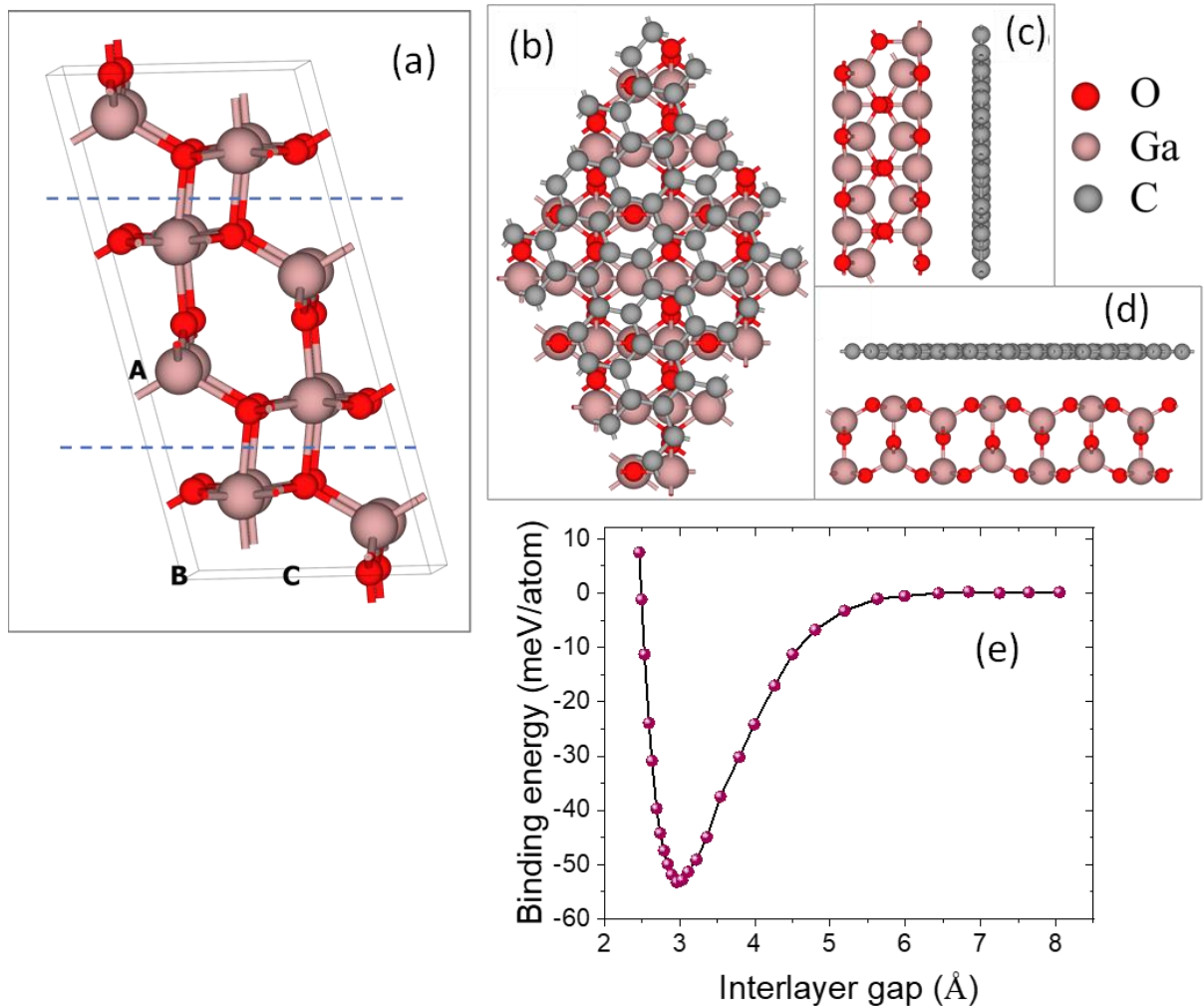

Figure 1, Atomic structure of the SLGO/graphene heterostructure and geometrical optimisation. (a) Monolayer $Ga_2O_3$ from a bulk cell. (b) Top view, (c) and (d) side view of the heterostructure. (e) Interlayer gaps vs. calculated binding energies.

The supercell of the bilayer structure has been constructed to a hexagonal lattice type containing 14 C, 8 Ga, and 12 O atoms with various lattice constants for different interlayer gaps. For each value of the gap, the supercell has been geometrically optimised using the generalized gradient approximation (GGA) with the para-metrization of Perdew-Burke-Ernzerhof (PBE), taking account of long range vdW interaction using semi-empirical corrections by Grimme DFT-D2 and counterpoise corrections. Density mesh cut-off was set to 75 Hartree, and k-points grid was set to 5 x 5 x 1. During the geometrical optimisation process, the force tolerance was 0.03 eV/Å and maximum stress was 0.001

eV/ Å³. Then the new values of gaps have been obtained after structural optimisation. At some interlayer gaps, e.g. some values smaller than 3 Å and between 3 – 5 Å, z-position of the atoms had to be fixed during the optimisation as the optimised gap values will fall into the 3 Å region if without z-axis anchoring.  The binding energy $E_b$ of each gap setting has been calculated using the numerical LCAO basis sets (linear combination of atomic orbitals) with hybrid functionals (HSE06) based on the formula $E_b = (E_t - E_{G\_s} - E_{SLGO\_s})/N_G$, where $E_t$ is the total energy of the vdW heterostructure, and $E_{G\_s}$ and $E_{SLGO\_s}$ are individual total energy of the strained graphene and SLGO respectively. $N_G$ is the number of graphene atoms in the cell (14). The calculated results are shown in Figure 1e. The calculated lowest binding energy is around -53.2 meV/atom at the interlayer separation around 3 Å, compared with the binding energy of around -45.5 meV/atom at around 3.6 Å separation for double layer graphene [28], and binding energy of around -70 meV/unit cell for graphene/hBN in reference [29]. Moreover, the calculated binding energy trend is similar to other types of bilayer heterostructures reported previously, e.g. in [28] [30].

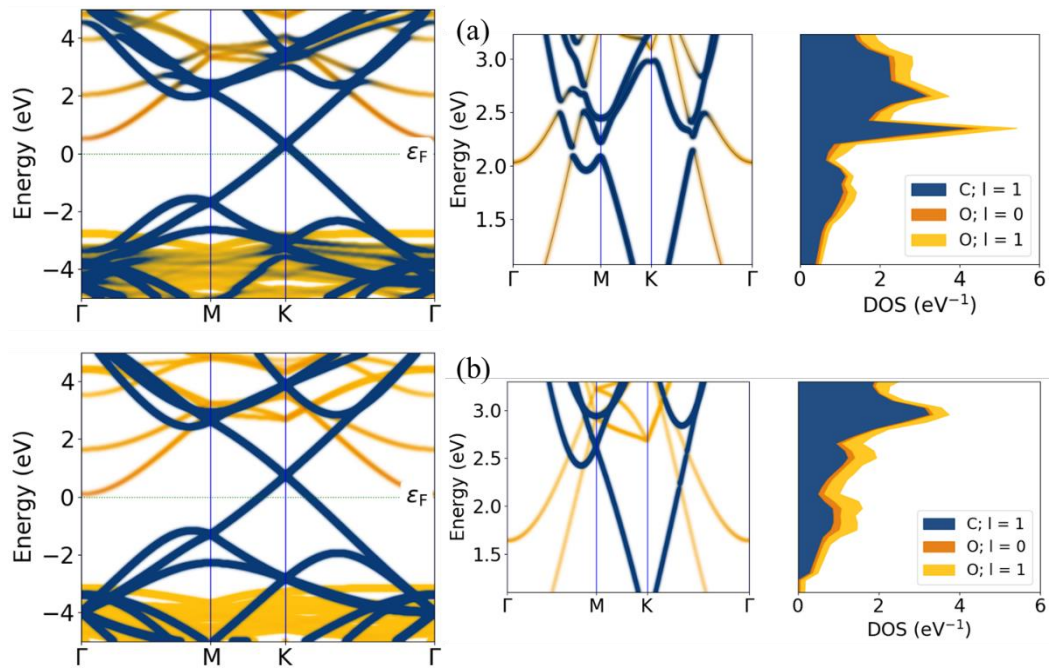

*Figure 2, Band structures for (a) small and (b) large interlayer distance. Close up view on the band opening region together with the projected density of states is shown on the right.*

Electronic band structure of this heterostructure has been calculated with the hybrid functionals that takes into account exact Fock exchange. The implementation was through HybridGGA HSE06 within the QuantumATK package. Fat band structures and projected density of state (PDOS) have been calculated, and the results are shown in Figure 2. Blue solid lines represent the electron densities at C 2p orbitals for the graphene, and yellow lines are for oxygen O 2p and O 2s. At around 2 eV in the conduction bands, it is clearly to see that for the narrow layer gaps (Figure 2a) the intersection of the graphene bands and $Ga_2O_3$ bands opens small band gaps in the Brillouin region of G-M, M point, and K-G, which may lead to interesting optoelectronics applications. Band opening near the Dirac cone of graphene/$C_3B$ ($C_3N$) heterostructures has been observed theoretically, which was explained by the stacking-order dependent charge transfer [31], where single layer $C_3B$ and $C_3N$ are all structurally similar to graphene [32]. It is the first time that this band opening effect has been observed for the heterostructure composed of wide-band semiconductor materials and graphene. For the large interlayer gaps the band splitting effect diminishes as shown in Figure 2b. The difference between two PDOSs of these two distinct interlayer gaps is a sharp rise at the band splitting region (around 2 eV) for the narrow interlayer gaps, which shows a localised density of states increase for the C 2p, correlating the band opening effect. Another explanation could be that the real, nonzero, coupling constant at the interface facilitates penetration of wave functions through the interface, and two wave functions (from both graphene and $Ga_2O_3$ bands) will hybridise [33]. This hybridisation will lift the degeneracy at cross points, resulting in band opening.

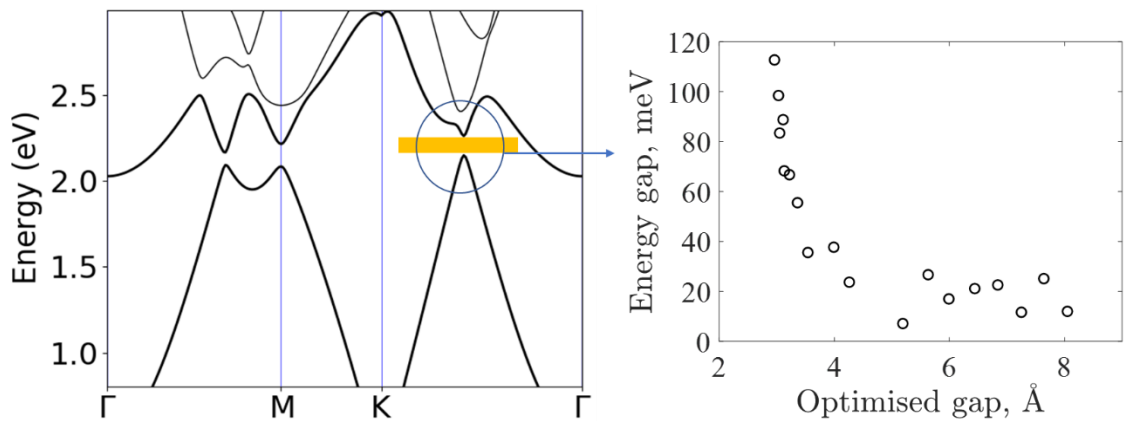

*Figure 3, Band opening at the point between K and G vs. optimised interlayer gap.*

Further calculations have been conducted to quantitively study the band opening effect for all interlayer gaps ranging from 2.96 Å to 8.05 Å (values after geometrical optimisation). As seen from Figure 3, one of split bands has been picked up for detailed investigation, which appears between K and G. The split bandgap is seen above 100 meV for very small interlayer distances due to electron density localisation on C 2p orbitals. The split bandgap abruptly drops as the interlayer distance increases until around 4 Å, after which the split bandgap stabilises at around 20 meV. This can also be explained qualitatively using one dimensional dispersion theory, where charge transfer induced small perturbation $\delta$ causes positive and negative parabolic band shape at the band opening point.

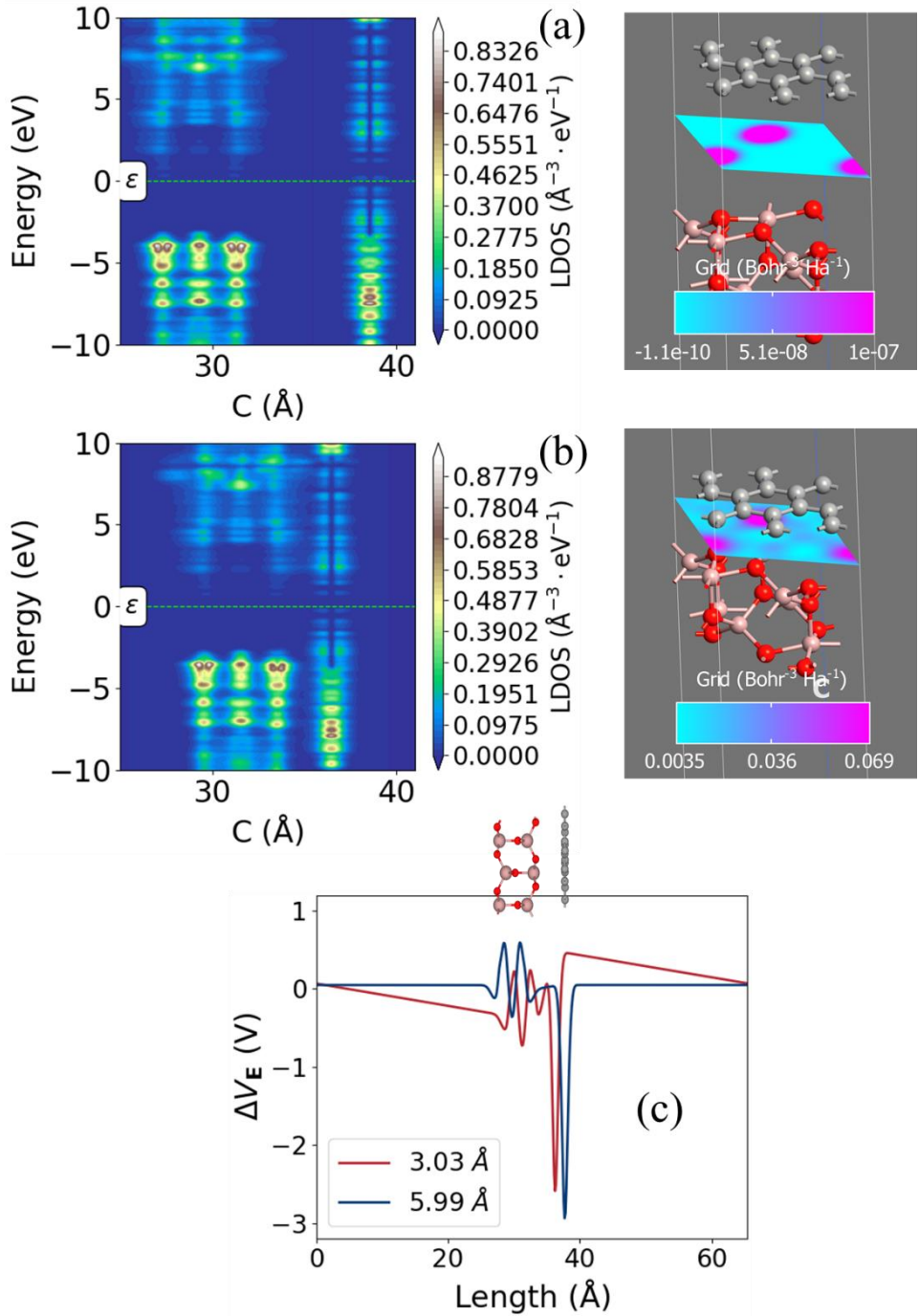

*Figure 4, Local density of states along c-axis for the heterostructure. (a) LDOS for large gap. (b) LDOS for small gap. (c) Electrostatic difference potentials for small and large interlayer gaps.*

Analysis of the local density of states (LDOS) along the c axis has been conducted and shown in Figure 4, where two types of the structures namely small interlayer gap (2.96 Å) and large interlayer gap (7.25 Å) have been demonstrated. Clearly with the large gap (Figure 4a), there is very weak interactions, hence very small DOS in the middle of two layers (maximum DOS at $1 \times 10^{-7}$ Bohr$^{-3}$Ha$^{-1}$). For the small interlay distance, the electron density of states in the middle of two layers is much larger (maximum around 0.069 Bohr$^{-3}$Ha$^{-1}$), which is 5 orders of magnitude higher. The intense interaction of the DOS of two layers is the possible cause for the band splitting, evidenced by the results shown in Figure 4b.

Electrostatic difference potential (EDP) for small and large interlayer gaps have been calculated. EDP represents the difference between the electrostatic potential of the self-consistent valence charge density and the electrostatic potential from a superposition of atomic valence densities. Figure 4c shows the 1D projected EDP along c-axis averaged over the two directions perpendicular to the c-axis, where the straight-line regions on both sides represent the vacuum regions on both sides of the hetero structure. For small interlayer gaps, charge transfer takes place at SLGO/graphene interface, causing the self-consistent electrostatic potential to differ from that of atomic valence densities. There is clearly a potential drop due to closely bonded interface. As opposed to the small interlayer gap, for the large interlayer gap (e.g., 5.99 Å in Figure 4c), there is nearly no charge transfer at the interface, leading to nearly zero voltage drops in the vacuum regions. It is the charge localisation at the interface that causes several interesting phenomena to occur, such as band-splitting in the conduction bands, and possible optical and mechanical switching effects.

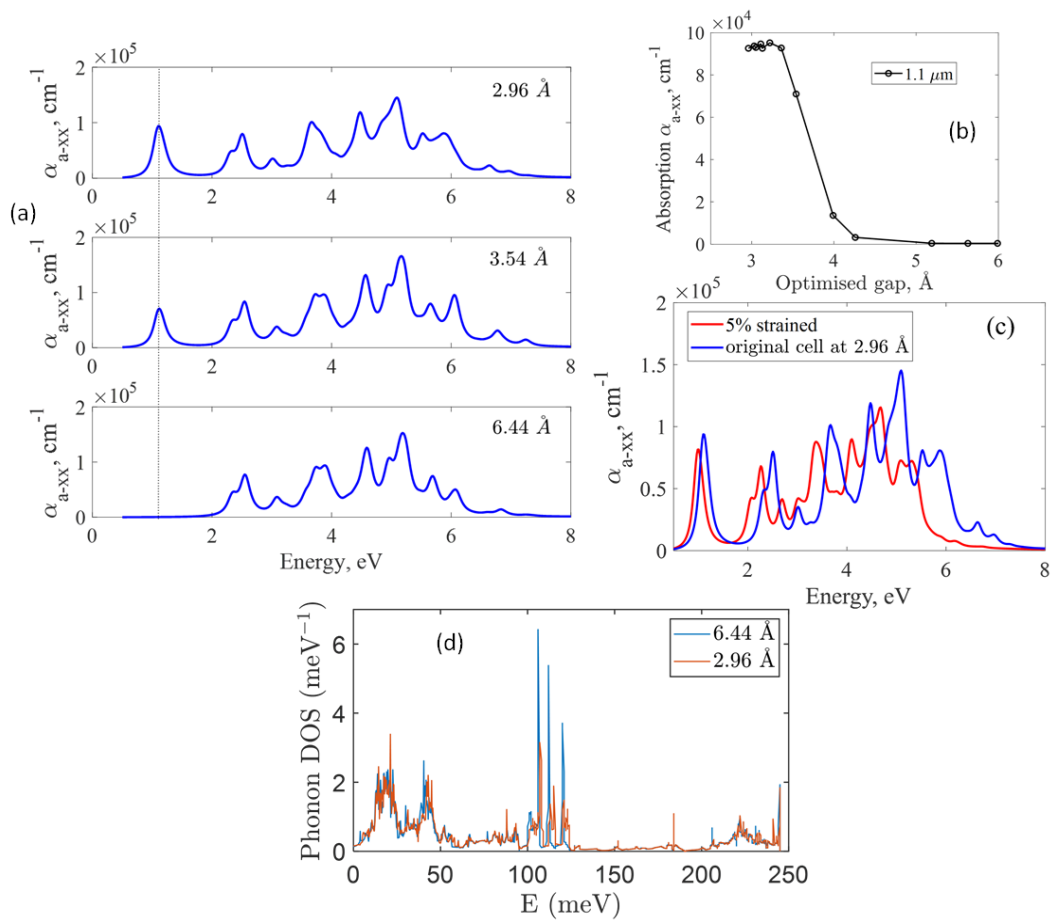

Figure 5, (a) Calculated in-plane optical absorption $\alpha_{a\text{-}xx}$ for small, medium and large interlayer gaps. (b) Peak amplitude of the optical absorption $\alpha_{a\text{-}xx}$ at 1.1 μm wavelength vs. optimised interlayer gaps.

*(c) The effect of tensile strain on the $α_{a\text{-}xx}$. (d) Calculated phonon density of states for small and large interlayer gaps.*

Optical absorptions for various interlayer gaps are calculated using the identical setup as calculating the bandgaps. Optical absorption ($α_a$) was calculated from complex part of the permittivity at a certain frequency $ω$, also known as the extinction coefficient ($κ$) using $α_a = \frac{2ωκ}{c}$. $κ$ has been derived from the susceptibility tensor. The complete description of the calculating procedure was in a previous paper [34]. In the optical absorption calculations, only the interlayers gaps that were geometrically optimised without anchoring the z-axis have been taken into considerations, i.e. the gaps near 3 Å, very sparse points between 3 - 5 Å and points greater than 5 Å. As the points residing in the steep slope regions in the binding energy graph tend to fall into the nearest shallow slope regions after structural optimisation without z-axis anchoring. The calculated results are shown in Figure 5a. In-plane absorptions in x-direction ($α_{a\text{-}xx}$) and y-direction ($α_{a\text{-}xx}$) are similar, hence only $α_{a\text{-}xx}$ is displayed. There is nearly zero absorption in z-direction, attributed to the atomic scale in the thickness direction of the monolayers. Interestingly, there is a peak at 1.1 eV (in infrared region) for small interlayer gaps (2.96 Å and 3.54 Å). This peak diminishes for large interlayer gap (6.44 Å). Calculation for a series of gaps have been made, and the results of absorption peak amplitudes at 1.1 eV versus optimised interlayer gaps are shown in Figure 5b, where a switching behaviour is demonstrated. For small gap values from 2.96 Å to 3.36 Å, absorption coefficients are almost at constant value, 9.3 x $10^4$ cm$^{-1}$. Absorption coefficient is switched to very low values (several orders of magnitude smaller than the so called 'on' state at small interlayer gaps) as the gap increases to around 4 Å. The optical absorption on-off ratio at 1.1 eV can reach up to 235, corresponding to 47.4 dB, which is approaching to the value reported in recent optical switching work [35]. This phenomenon can be applied as an optical switch, gated by the interlayer distance. Also, this switching effect can be utilised to realise an atomic level high-sensitivity strain sensor.

More analysis has been carried out to study mechanical strain effect on the optical absorption of the SLGO/graphene heterostructure. Shown in Figure 5c, the absorption has demonstrated a red-shift and slightly reduced amplitude for 5% strain in the a-direction of this hexagonal cell (at 2.96 Å interlayer gap), which has potential in flexible or wearable device applications.

Phonon density of states for the small and large interlayer gaps have been calculated using the semi-empirical calculator with Slater-Koster parameters, and result is shown in Figure 5d. As the gap gets small, the structure gets softened as compared to the large gap structure. This is interpreted that individual Ga$_2$O$_3$ and graphene displays high elastic modulus, however when they are bonded together using vdW force, the overall structure is softened, as the interlayer interaction reduces the elastic modulus of the whole heterostructure reduces.

Optical absorptions of single layer Ga$_2$O$_3$ and graphene have been calculated and shown in the new Figure 6. Combined with results shown in Figure 5, the interaction between the two layers mainly affects the absorption spectrum in the low energy region, which is determined by the intraband transition of the graphene.

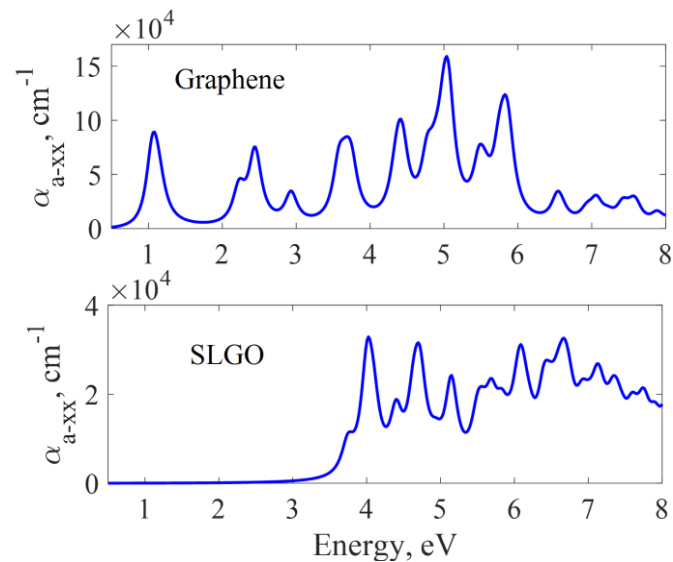

*Figure 6, Calculated optical absorption for single layer Ga$_2$O$_3$ and graphene.*

**Conclusion**: In summary, a novel heterostructure constructed from ultra-wide band semiconductor material – Ga$_2$O$_3$ and graphene has been investigated by using DFT with various interlayer distances. Through geometrical optimisation, it has been found that the binding energy of this heterostructure has demonstrated a similar trend as in other graphene-based bilayer structures. Electronic and optical properties of the heterostructure have been calculated, and band opening for small gaps was unveiled. On-off switching behaviour of the in-plane optical absorption coefficient at 1.1 eV has also been demonstrated. These new findings are likely originated from the interaction of heavy electron effective mass of wide bandgap semiconductor and very light effective mass of graphene. This work paves a way for the future tuneable optoelectronic devices based on adjusting the interlayer distance of the heterostructures.

**Acknowledgement**:

The author would like to thank …

**Declaration of Competing Interest:**

The author declares that they have no known competing financial interests or personal relationships that could have appeared to influence the work reported in this paper.

Graphic abstract:

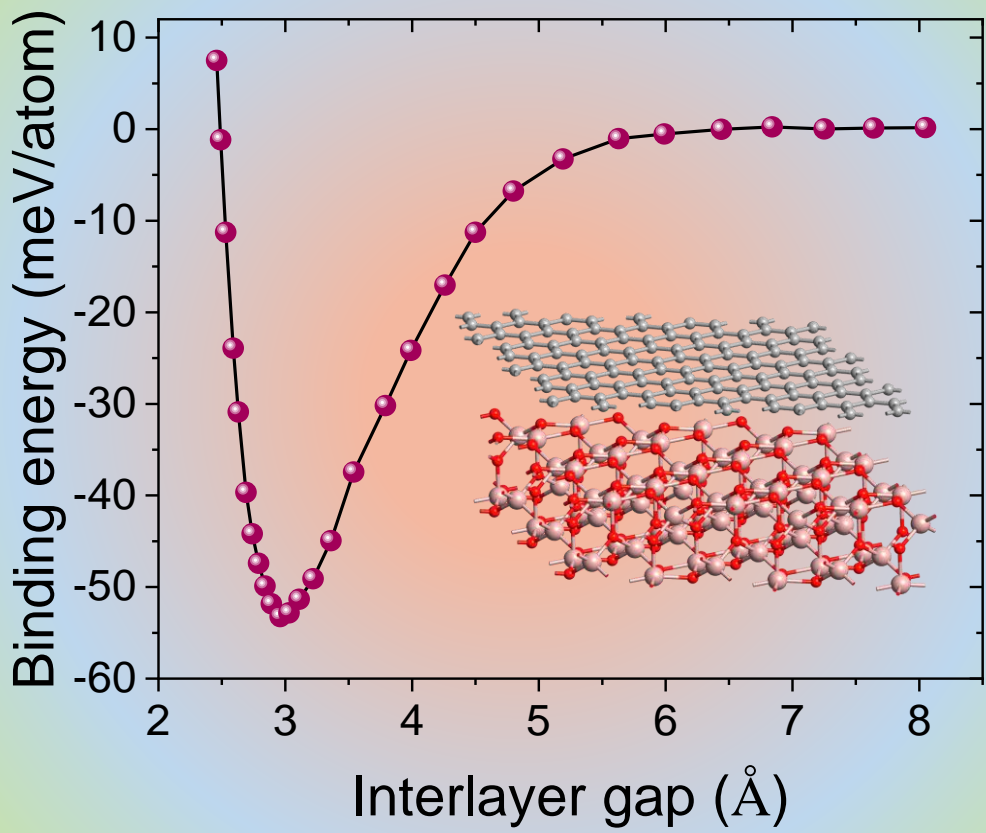